# Synchrony parameter dependent transformation equations and some of their particular values

Kinematics

Bernhard Rothenstein
1) Politehnica University of Timisoara, Physics Department,
Timisoara, Romania brothenstein@gmail.com

**Abstract.** We show that alternative relativity theories that are essentially based on varied distant clock synchronization procedures can be recovered by using the standard Lorentz-Einstein transformations for the space-time coordinates of the same event. Through this approach we offer modest support for the Rizzi et. alt. stating that: "Once correctly and explicitly phrased, the principles of special relativity theory allow for a wide range of **theories** that differ from the standard Einstein's theory only for the difference in the chosen synchronization procedure, but are wholly equivalent to special relativity theory in predicting empirical facts". Our approach requires of the reader no more then a correct understanding of the physics behind the Lorentz-Einstein transformations equations which we use.

**1. Introduction**

Einstein's special relativity theory (**SRT**) [1] and Reichenbach's relativity theory (**RRT**) [2] agree in the way in which they state the
**principle of relativity**
-The existence and equivalence of inertial frames (principle of relativity)
In (**SRT**) the second postulate is stated as:
-The constancy of the **one-way** speed of light in empty space relative to these frames (the principle of the constancy of the of the one way speed of light),
whereas in RRT it is stated as
-The constancy of the **two-way (round trip)** speed of light relative to these frames (the principle of the constancy of the round trip speed of light.

The synchronization of distant clocks involves signals that propagate between them, and so we are obliged to make a net distinction between their synchronization using an one-way synchronization procedure (standard clock synchronization proposed by Einstein [1]) and the two-way clock synchronization proposed by Reichenbach [2]. The difference between them is illustrated in Figure 1 on a classical space-time diagram.

Both theories involve the inertial reference frames I and I' positioned in the so-called **standard arrangement**, which leads to the simplest results. The corresponding axes of the two frames are parallel to each other, at the origin of time in the two frames (t=t') their origins O and O' respectively are located at the same point in space and I' moves with constant speed V relative to I in the positive direction of the permanently overlapped OX(O'X') axes.

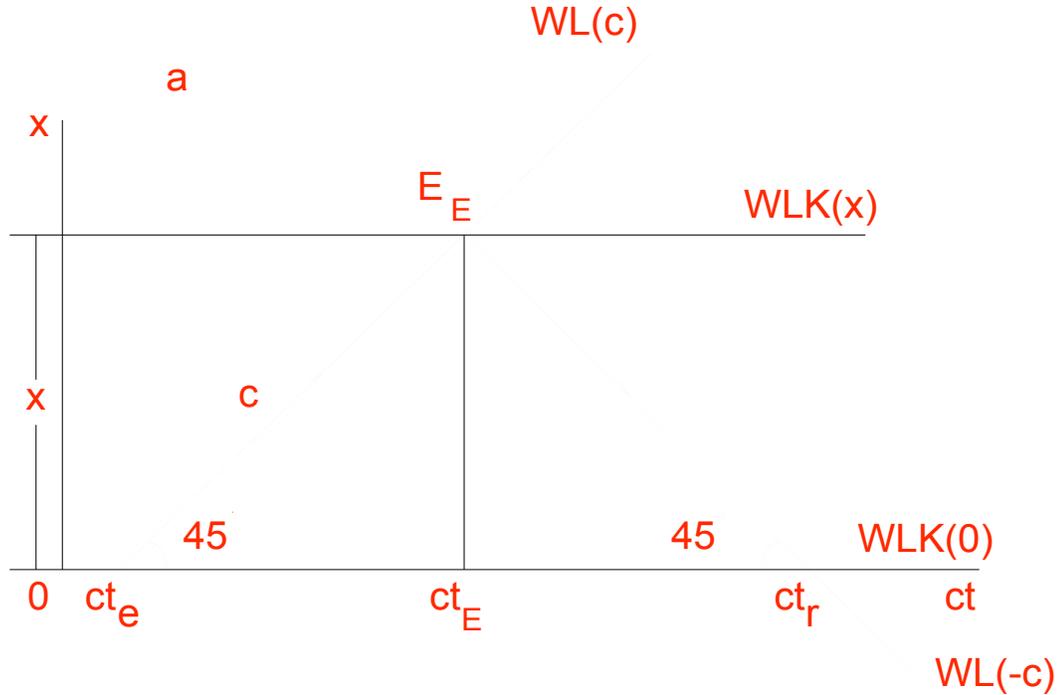

*Figure 1.a. Illustrating the standard synchronization of the clocks K(0) and K(x) of the I frame on a classical space-time diagram*

Figure 1a illustrates the standard (Einstein) synchronization of the clocks $K_0(0)$ located at the origin O of I and $K(x)$ located at a distance x apart from it. $WLK_0(0)$ represents the world line of clock $K_0(0)$, i.e the geometric locus of the successive events it generates at different times. When clock $K_0(0)$ reads $t_e$ a source of light $S(0)$ located at the origin O emits a light signal (WLc) in the positive direction of the OX axis. This light signal arrives at the location of clock $K(x)$ when it reads $t_E$, generating the event $E_E(x, t_E)$. The synchronizing signal is instantly reflected back (WL-c) and is received at the origin O when clock $K_0(0)$ reads $t_r$. The geometry of the space-time diagram tells us that

$$ct_E = ct_e + \frac{x}{c} \qquad (1)$$

$$ct_E = ct_r - \frac{x}{c}. \qquad (2)$$

Eliminating x between (1) and (2), we obtain

$$t_E = \frac{1}{2}(t_r + t_e) \qquad (3)$$

Given this result and knowing the readings of his wrist watch $K_0(0)$ when the light signal is emitted ($t_e$) and received back after reflection ($t_r$), the observer $R_0(0)$ who handles the source of light is able to assign to clock $K(x)$ the time $t_E$. **We emphasize that in the standard synchronization procedure it is assumed that the synchronizing signal propagates forward and backward with the same invariant speed.**

Figure 1b illustrates the synchronization of clocks $K(0)$ and $K(x)$ in (**RRT**). The light signal, emitted at a time $t_e$ and propagating with speed $c_+<c$, arrives at the location of clock $C(x)$ when it reads $t_R$, propagates back, after reflection with speed $c_-$ and is finally received back when clock $K_0(0)$ reads $t_r$. The geometry of the space-time diagram tells us that

$$t_R = t_e + \frac{x}{c_+} \qquad (4)$$

$$t_R = t_r - \frac{x}{c_-}. \tag{5}$$

and, as we can see, $c_- > c$. Making the assumption that

$$c_+ = \frac{c}{n} \tag{6}$$

where the synchrony parameter n>1 and taking into account that the round trip speed of light c is defined by [2] as

$$\frac{2}{c} = \frac{1}{c_+} + \frac{1}{c_-} \tag{7}$$

we obtain that

$$c_- = \frac{c}{2-n}. \tag{8}$$

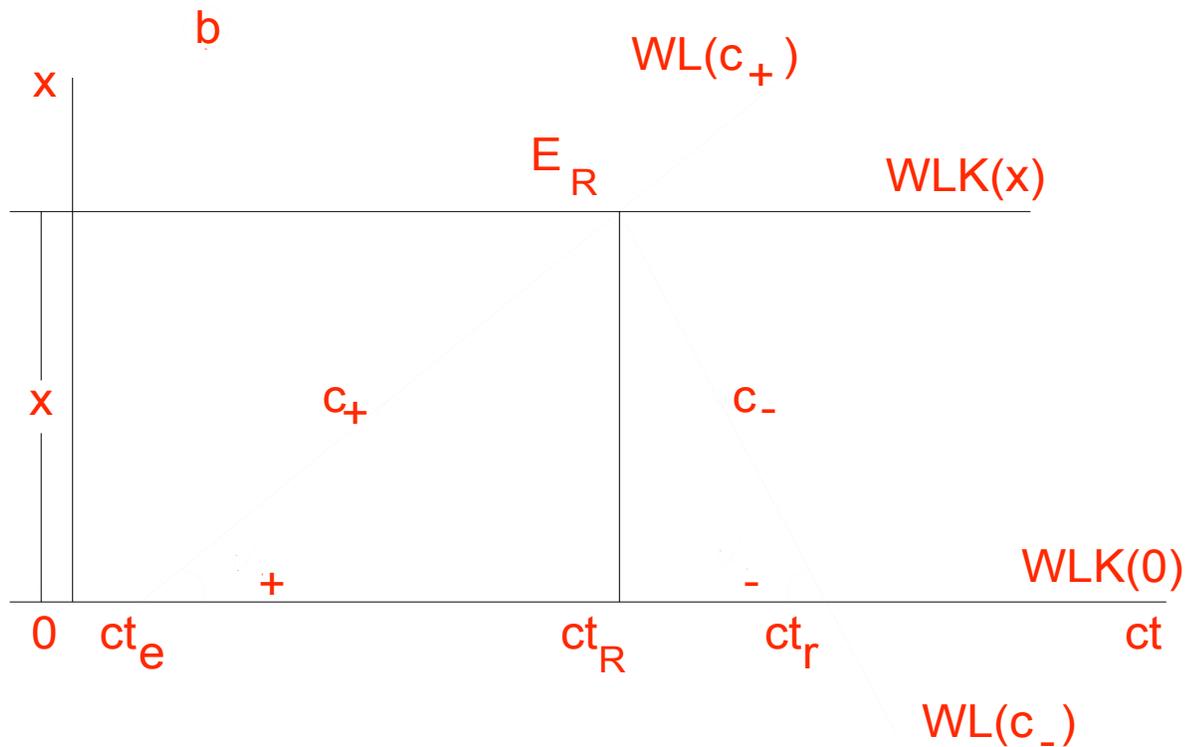

*Figure 1.b Illustrating the Reichenbach clock synchronization of clocks $K_0(0)$ and $K(x)$ of the I frame .on a classical space-time diagram*

With n=1 we recover (3). By eliminating x between (4) and (5), we obtain that the clock synchronized a la Reichenbach displays the time

$$t_R = t_e + \frac{n}{2}(t_r - t_e). \tag{9}$$

As we can see, different values of n and the same values $t_e$ and $t_r$ lead to different values of $t_R$. Within the limits of (**SRT**), which we do not intend to leave, we impose the condition n>1; c is an upper limit for the speed at which particles could move.

The standard clock synchronization is under plausible critique. The reasons are motivated by Ungar [3] and Ohanesian [4] as follows:

*"The standard Einstein synchronization implicitly assumes equal one-way speeds for the outward and return signals, whereas the nonstandard Reichenbach synchronization implicitly assumes unequal one-way speeds for the outward and return signals. It is impossible to decide which of these assumptions is correct, because synchronization and the one-way speed of light are joined in a **vicious circle, each depends on the other.** To determine the one-way speed of light experimentally, we need to measure the "time of flight" of a light signal over a given distance, say, the distance between clocks A and B. If the clocks have been synchronized with light signals, then **such a measurement is logically circular.**"*

**2. Formalism to deal with Reichenbach's relativity theory or synchrony parameter (n) dependent transformation equations**

By involving the concept of round trip speed of light, transformation equations for the space-time coordinates which include the synchrony parameter n are derived The derivations avoid making use of the Lorentz-Einstein transformations. Presented in their standard shape, the Lorentz-Einstein transformations for the space time coordinates of the same event detected form I and I', respectively, are

$$x = \gamma(x' - Vt'_E) \quad (9) \qquad\qquad t_E = \gamma\left(t'_E + \frac{V}{c^2}x'\right) \quad (10)$$

and their inverse are

$$x' = \gamma(x + Vt_E) \qquad (10) \qquad\qquad t'_E = \gamma\left(t_E - \frac{V}{c^2}x\right) \qquad (11)$$

where $\gamma = \sqrt{1 - \frac{V^2}{c^2}}$ represents the Lorentz factor. The Lorentz-Einstein transformations are a direct consequence of the fact that the clocks of the inertial reference frames involved are standard synchronized.

Ungar [3], Edwards [5] Mansouri and Sexl [6] take the concept of two-way speed of light seriously and, therefore, derive n dependent transformation equations which are the result of the fact that $c_+=c/n$ with n>1. Other authors derive such transformations which are, as we will show, the result of a particular value of n. Thus, for n=1+V/c we recover the transformation equations proposed by Selleri [7], Tangherlini [8] and Abreu and Guerra [9].

The Lorentz-Einstein transformations (**LET**) play a fundamental part in our derivations. We involve them via the following statements:

-The **(LET)** hold exactly only in the case when the times involved are the readings of standard synchronized clocks in all the inertial reference frames involved

The two statements lead to n dependent transformation equations. Well defined values of n recover the transformation equations that account for the situations in which in one of the involved inertial reference frames the clocks are standard synchronized whereas in the other the synchronization is performed by a subluminal signal propagating with speed $c_+=c/n$ where n>1.

-In each of the inertial reference frames involved, the clocks could be synchronized using a signal that propagates with the invariant speed c, but could also be synchronized using a signal that propagates with a subluminal signal $c_+=c/n$ where n>1; we must remember to take into account that, within the limits of **(SRT)**, c is an upper limit for the speed at which objects could move.

Consider the identical clocks $K'_0(0); K'_1(x'); K'_2(x')$ of the I' frame. The first is located at the origin O', while the other two are located at the same point M'(x'). Clocks $K'_0(0)$ and $K'_1(x')$ are synchronized using a signal that propagates with the invariant speed c and is emitted when the clocks of I' read t'=0. When the synchronizing signal arrives at the point M'(x'), the clock $C'_1(x')$ reads $t'_E$. If

$$t'_E = \frac{x'}{c} \qquad (12)$$

then we say that the two clocks involved are standard synchronized. The clocks $K'_0(0); K'_2(x')$ are synchronized using a signal that propagates with the subluminal speed $c'_+ = \dfrac{c}{n}$ that is emitted from the origin O' when the clocks of I' read t'=0. It arrives at the location of the clock $C'_2(x')$ when it reads

$$t'_n = \frac{nx'}{c} \tag{13}$$

and we say that the two clocks are nonstandard synchronized using a subluminal signal. Equations (12) and (13) lead to the following relationship between the readings of the two clocks,

$$t'_E = t'_n + (1-n)\frac{x'}{c} \tag{14}$$

a fundamental equation in our derivations. When it arrives at the location of clocks $K'_1(x'); K'_2(x')$, the subluminal signal generates the event

$$E'[x'; t'_E = t'_n + (1-n)\frac{x'}{c}].$$

Performing the (**LET**) of the space-time coordinates of event E' we obtain the n dependent transformation equations

$$x = \frac{\left[1 + \dfrac{V}{c}(1-n)\right]x' + Vt'_n}{\sqrt{1 - \dfrac{V^2}{c^2}}} \tag{15}$$

and

$$t_E = \frac{t'_n + \left(1 - n + \dfrac{V}{c}\right)\dfrac{x'}{c}}{\sqrt{1 - \dfrac{V^2}{c^2}}}. \tag{16}$$

Simple algebra leads to the following inverse transformations

$$x' = \frac{x - Vt_E}{\sqrt{1 - \dfrac{V^2}{c^2}}} \tag{17}$$

and

$$t'_n = \frac{\left[1 + \dfrac{V}{c}(1-n)\right]t_E - \left(1 - n + \dfrac{V}{c}\right)\dfrac{x}{c}}{\sqrt{1 - \dfrac{V^2}{c^2}}}. \tag{18}$$

As expected, for n=1 we recover the (**LET**). Different values of the synchrony parameter n>1 lead to different theories reviewed in the following chapters. We emphasize that the equations derived above avoided the concept of the two-way speed of light and worked only with concepts with which (**SRT**) agrees. As we can see, (17) is n independent and so it should have the same shape in all theories that correspond to different values of n.

## 3. Absolute simultaneity: Selleri [7], Tangherlini [8] and Abreu and Guerra [9].

We impose the condition that the transformation equations for the time coordinates do not depend on the space coordinates. Under such conditions, (18) tells us that

$$1 - n + \frac{V}{c} = 0 \tag{19}$$

i.e.

$$n = 1 + \frac{V}{c} \tag{20}$$

The subluminal synchronizing signal propagates in the positive direction of the overlapped OX(O'X') axes with speed

$$c'_+ = \frac{c}{1 + \frac{V}{c}}. \tag{21}$$

With n=1+V/c (17) and (18) become

$$x' = \frac{x - Vt_E}{\sqrt{1 - \frac{V^2}{c^2}}} \tag{22}$$

$$t'_n = \sqrt{1 - \frac{V^2}{c^2}} t_E \tag{23}$$

Thus we have recovered the results obtained by Selleri (7) and Tangherlini [8] based on the concept of two-way speed of light and by Abreu and Guerra [9] based on length contraction and on the external clock synchronization. With n=1+V/c (15) and (16) become

$$x = \sqrt{1 - \frac{V^2}{c^2}} x' + \frac{Vt'_n}{\sqrt{1 - \frac{V^2}{c^2}}}. \tag{24}$$

and

$$t_E = \gamma t'_n.$$

**Conclusions**

By consecutively applying the Lorentz-Einstein transformations we have recovered all the results offered by alternative relativity "theories" that are based on special nonstandard clock synchronization procedures, such as the one derived by Selleri or the one revealed by Abreu and Guerra without using the Lorentz-Einstein transformations. The new approach that we present in this paper emphasizes the fact that it is very usefully to define the clocks that display the time coordinates involved in these synchronization procedures. A hint from Mermin [10] applies well in this context: *"Take a situation which you don't fully understand. Find a new frame of reference in which you do understand it. Examine it in that new frame of reference. Then translate your understanding in that new frame back into the language of the old one."* The author feels that Einstein's special relativity theory is mighty enough that the derivation of other non-rival theories can be avoided in the limits of the studied problem.